\documentclass[reprint,amssymb,amsmath,aps,pra,superscriptaddress]{revtex4-1}

\usepackage{graphicx}
\usepackage{dcolumn}
\usepackage{bm}
\usepackage{amsmath}
\usepackage{amssymb}
\usepackage{natbib}

\begin{document}

\title{Manipulation of gravitational quantum states of a bouncing neutron with the GRANIT spectrometer}

\author{B.~Clément}\email[corresponding author : ]{bclement@lpsc.in2p3.fr}
\affiliation{LPSC, Universit\'e Grenoble-Alpes, CNRS/IN2P3, 
53, avenue des Martyrs, Grenoble, France}
\author{S.~Bae\ss{}ler}
\affiliation{Physics Department, University of Virginia, 382 McCormick Road, Charlottesville, Virginia 22904, USA}
\affiliation{Oak Ridge National Laboratory, Bethel Valley Road, Oak Ridge, TN 37831, USA}
\author{V.~V.~Nesvizhevsky}
\affiliation{ILL, Institut Laue Langevin, 
71, avenue des Martyrs, Grenoble, France}
\author{E.~Perry}
\affiliation{ILL, Institut Laue Langevin, 
71, avenue des Martyrs, Grenoble, France}
\affiliation{University College London, Department of Physics \& Astronomy, Gower St, London, WC1E 6BT}
\author{G.~Pignol}
\affiliation{LPSC, Universit\'e Grenoble-Alpes, CNRS/IN2P3, 53, avenue des Martyrs, Grenoble, France}
\author{J.A.~Pioquinto}
\affiliation{Physics Department, University of Virginia, 382 McCormick Road, Charlottesville, Virginia 22904, USA}
\author{K.~V.~Protasov}%
\affiliation{LPSC, Universit\'e Grenoble-Alpes, CNRS/IN2P3, 53, avenue des Martyrs, Grenoble, France}
\author{D.~Rebreyend}
\affiliation{LPSC, Universit\'e Grenoble-Alpes, CNRS/IN2P3, 53, avenue des Martyrs, Grenoble, France}
\author{D.~Roulier}
\affiliation{ILL, Institut Laue Langevin, 
71, avenue des Martyrs, Grenoble, France}
\author{L.~Shen}
\affiliation{Physics Department, University of Virginia, 382 McCormick Road, Charlottesville, Virginia 22904, USA}
\affiliation{Department of Physics, University of Washington, Seattle WA, USA}
\author{A.~V.~Strelkov}
\affiliation{Joint Institute for Nuclear Research, RU-141980 Dubna, Russia}
\author{F.~Vezzu}
\affiliation{LPSC, Universit\'e Grenoble-Alpes, CNRS/IN2P3, 53, avenue des Martyrs, Grenoble, France}


\begin{abstract}
The bouncing neutron is one of the rare system where gravity can be studied in a quantum framework. To this end it is crucial to be able to select some specific gravitational quantum state (GQS). The GRANIT apparatus is the first physics experiment connected to a superthermal helium UCN source. We report on the methods developed for this instrument showing how specific GQS can be favored using a step between mirrors and an absorbing slit. We explore the increase of GQS separation efficiency by increasing the absorber roughness amplitude, and find it is feasible but requires a high adjustment precision. We also quantify the transmission of the absorbing slit leading to a measurement of the spatial extension of the neutron vertical wave function  $z_0 = \hbar^{2/3}\left(2m^2g\right)^{-1/3} = 5.9\pm0.3\,\mu$m.

\end{abstract}

\maketitle

\section{\label{intro}Introduction}

We consider the problem of a neutron with negligible vertical velocity bouncing on a flat mirror. The vertical motion is that of a particle trapped in a gravitational potential well, infinite at mirror surface $z=0$ and $V=mgz$ above the mirror. The stationary Schr\"odinger equation describing the vertical motion reduces to an Airy equation:
\begin{equation}
\frac{d^2\psi(Z)}{dZ^2} =(Z-\varepsilon_i)\psi(Z),
\end{equation}
where $Z = z/z_0$ with $z_0 =\hbar^{2/3}\left(2m^2g\right)^{-1/3} \approx 5.87\,\mu$m.
Boundary conditions $\psi(0)=0$ and $\psi(+\infty)=0$ impose the quantization of the energy levels, which are given for the $i$-th state by $E_i=mgz_0\varepsilon_i\approx 0.602 \varepsilon_i$\,peV, where $-\varepsilon_i$ is the $i$-th zero of the Airy $Ai$ function. The corresponding wave functions are:
\begin{equation}
\label{eqn:schr}
\psi_i(z) = \frac{1}{z_0|Ai'(-\varepsilon_i)|}Ai\left(\frac{z}{z_0}-\varepsilon_i\right).
\end{equation}
The $z_0$ parameter characterizes the spatial extension of the wave functions. 
Ultra-cold neutrons (UCN)~\cite{Lushchikov69,Steyerl69}, with kinetic energies smaller than 250\,neV can be used to study gravity in this quantum context. The first evidence of gravitational quantum states (GQS) was observed in the transmission of an absorbing slit~\cite{Nesvizhevsky2000,Nesvizhevsky2002}. More recently devices have been devised to study the transition between GQS such as the qBounce~\cite{Jenke2011,Cronenberg2018} and GRANIT~\cite{Roulier2015,Baessler2011} experiments at Institut Laue-Langevin. The spatial shape of the wave function has also been observed using a magnifying glass rod~\cite{Ichikawa14}.
Such studies necessitate the ability to select GQS. This paper presents the methods developed  for the GRANIT experiment using a step between mirrors and an absorbing slit to perform this selection and measure the $z_0$ parameter.

\section{\label{experiment}Experimental setup}
The GRANIT installation at ILL consists of two coupled apparatuses: a cryogenic UCN source~\cite{Schmidt-Wellenburg09,Zimmer11} and a gravitational level spectrometer~\cite{Roulier2015}. The UCN source relies on inelastic scattering of cold neutrons at $\lambda=0.89$\,nm on phonons in superfluid helium to produce UCN~\cite{Golub75}. The interior of the source and the extraction connecting the source to the spectrometer are covered with PFPE grease, providing a high reflection coefficient but limiting the maximum velocity of the neutrons to 4.6\,m.s$^{-1}$. GRANIT is the first physics experiment connected to this kind of superthermal helium UCN source.

The source produces a continuous flux of about 2000 UCN\,s$^{-1}$.
Neutrons are stored in a cylindrical copper volume where many bounces randomize their trajectories. Neutrons with vertical velocities smaller than $0.05\,$m.s$^{-1}$ can escape through a 10\,cm long, 30\,cm wide and 127\,$\mu$m high semi-diffusive slit~\cite{Psw07,Barnard08} with a rate of approximately 1 UCN per second. This slit is composed of two DLC coated glass mirrors, a smooth bottom mirror and a diffusive top one, separated by three small mylar wedges. 
Neutrons getting out of the slit propagate over a 25\,cm glass mirror with a mean roughness of 0.5\,nm and a planarity of 80\,nm. 

\begin{figure}[!htb]
\includegraphics[width=0.45\textwidth]{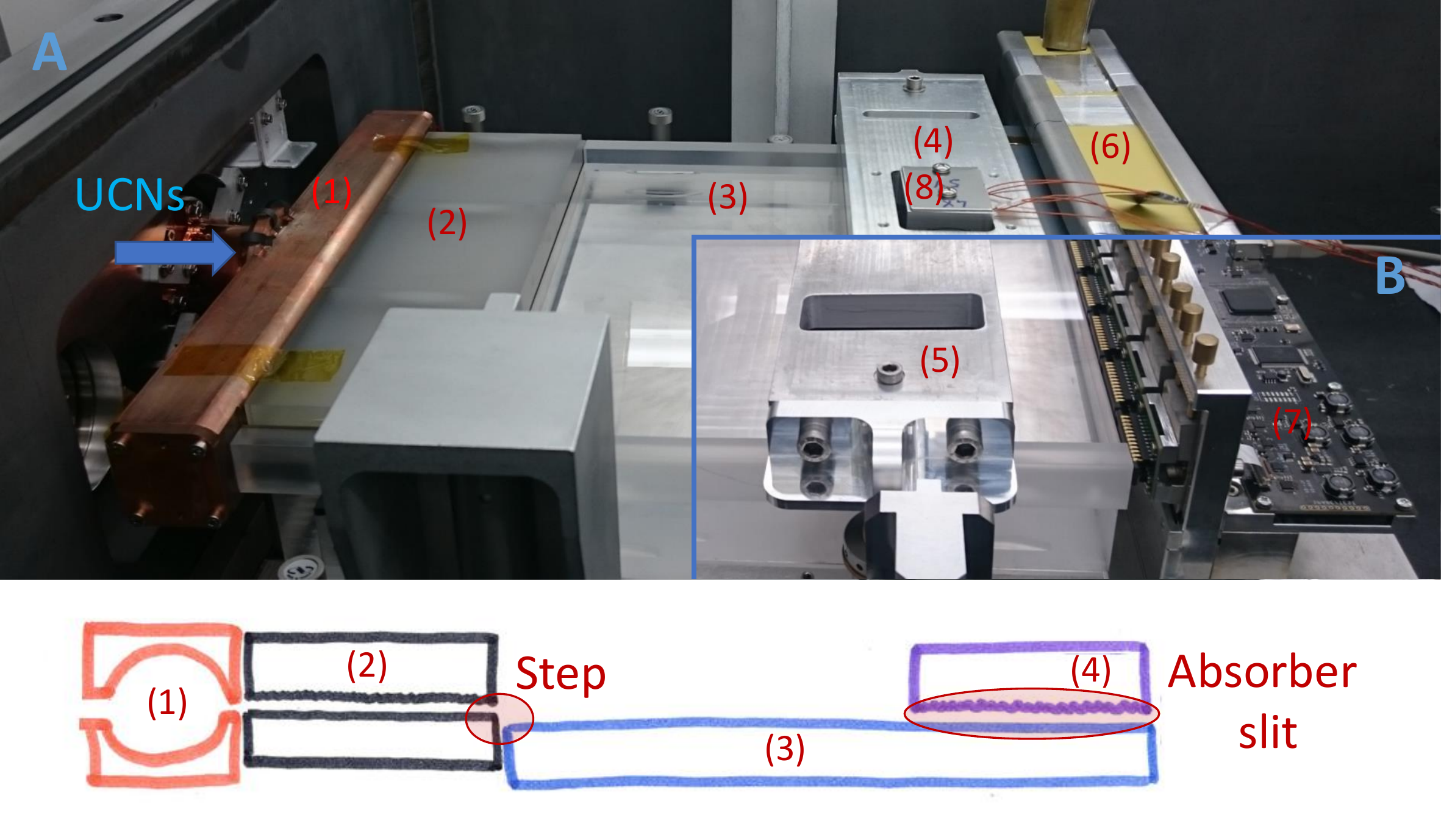}
\caption{\label{fig:granitspectro}. The GRANIT spectrometer in absorber transmission (main image A) and its modification for the wave function (sub-image B) configuration. The position of the step and absorber slit are detailed on the bottom drawing (not to scale).
From left to right, the copper randomization volume (1), the semi-diffusive slit (2), the propagation mirror (3) with the settable absorber (at the end (4)  or in the center (5) ) and the detectors ($^3$He counter in A (6), $^{10}$B coated charged coupled devices (CCD) $\textrm{UCNBoX}$ (7) in B). On the top of the absorber one can see the dual axis inclinometer (8) used in the setting of the parallelism.}
\end{figure}

The spectrometer, shown in Fig.~\ref{fig:granitspectro}, features two systems designed to suppress low energy and high energy states.
First, at the transition between the semi-diffusive slit and the main mirror, a step can be set. When falling over the step of height $s$ the lower quantum states scatter to higher states with a probability $p_{i\rightarrow j}=\left|\int \psi_i^*(h)\psi_j(h+s)\textrm{d}h\right|^2$, effectively suppressing up to the $i$-th state if $s\gtrsim z_0\epsilon_i$~\cite{Nesvizhevsky04}. 
To eliminate high energy states, an absorber slit is placed either at the end or in the center of the mirror. It consists of a block of glass with a rough chemically etched surface coated with gadolinium. The roughness is approximately $6\,\mu$m~\cite{Escobar15}, close to $z_0$. 
It is chosen to be sufficiently higher than those in any previous experiment with GQS in order to increase the efficiency of this device. The absorber is placed on top of the propagation mirror, parallel to its surface. Depending on the height of the slit, high energy states have a larger presence probability on the absorber and scatter into higher states and random direction. Scattered neutrons are then absorbed by the gadolinium or are lost through the mirror. The physics of the rough absorber is described in~\cite{Voronin2006,Meyerovich06,Meyerovich07,Escobar11}. 

The setting of the step is ensured by moving the propagation mirror with respect to the slit bottom mirror using three micrometric screws. Two Keyence LK-G5000 laser height sensors, on each side of the spectrometer are used to control both the height difference before and after the step and the relative parallelism, moving the sensor along an arc around a fixed axis. 
The precision of the measurement is better than a micron. The precision of the setting itself is estimated by its reproducibility. By evacuating the chamber to $10^{-5}$\,mbar then back to atmospheric pressure and remeasuring the step we noticed variations of a few microns due to small displacement of parts of the apparatus. For data presented in section~\ref{sec:ccd} this reproducibility error is around 6\,$\mu$m (1\,$\sigma$). The support of the mirrors was later modified to ensure a better reproducibility of 2\,$\mu$m.


A set of three piezoelectric actuators allows to remotely modify the height of the slit. To align the absorber, it is placed on three 127\,$\mu$m mylar wedges. This position is considered to be $h=127\,\mu$m with surfaces parallel. Due to the roughness of the absorber surface the parallelism cannot be guaranteed to be much better than $70\,\mu$rad ($6\,\mu$m over the $9$\,cm length). The absorber is then moved up, the wedges removed, and the absorber lowered so the actuators allow a movement from $20$ to $70\,\mu$m above the main mirror. The two laser sensors are used to check the height of absorber top surface relative to the $127\,\mu$m reference and a dual axis inclinometer ensures the parallelism is not degraded any further than by a few $\mu$rad. The precision on a variation of height is $1\,\mu$m over the $50\,\mu$m range. The reproducibility of this alignment is checked  in a similar manner as the step leading to a $1\,\mu$m accuracy (1\,$\sigma$).

\section{\label{sec:ccd}Manipulation of quantum states}

The detail studies of the properties of the bouncing neutron quantum states implies that one can select one or several quantum states. Discriminating between the contributions of each state requires access to the vertical distribution of the neutrons. To this end we use a dedicated position sensitive detector composed of 8 CCD sensors~\cite{Bourrion2018} coated with $^{10}$B~\cite{Clement19} with a spatial resolution of $\sigma_d=2\,\mu$m~\cite{Clement22}.
The absorber is placed in the center of the mirror to allow free bounces of UCNs on the mirror after getting out of the absorber slit. Before reaching the detector surface, UCN undergo a free fall over a horizontal distance of $x=1.0\pm0.1\,$mm due to the casing of the CCD sensor. 
This setup is presented in Fig.~\ref{fig:granitspectro}B.

\begin{figure}[!htb]
\includegraphics[width=0.45\textwidth]{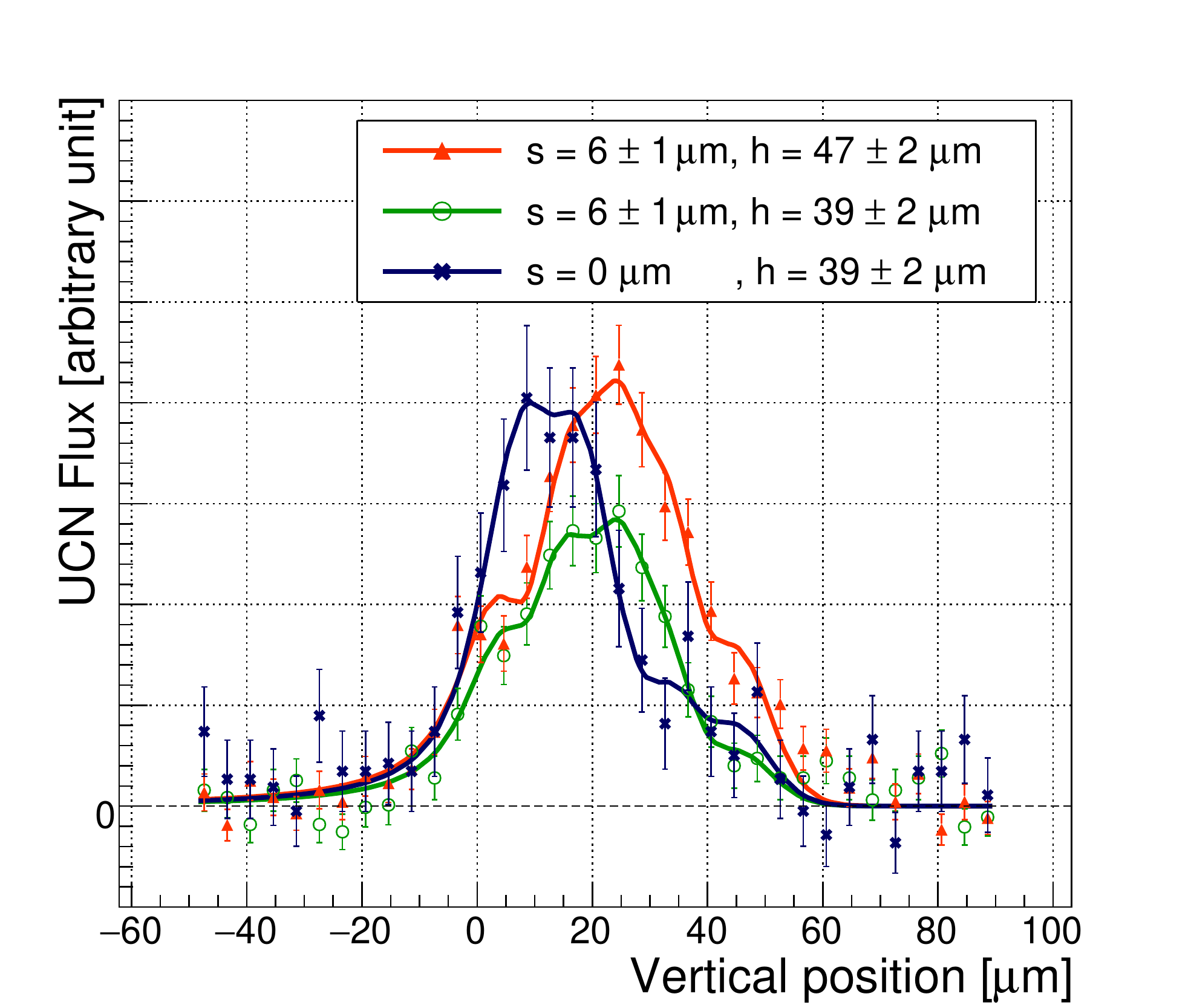}
\caption{\label{fig:plotccd} Vertical distribution of detected UCN for the three studied configuration: low absorber without step (blue crosses), low absorber with step (green circles) and high absorber with step (red triangles).}
\end{figure}

The relative alignment of the sensors is done by reconstructing from data the position and orientation of the slit center (combining all data sets) fitting the data to a Gaussian profile. 
The data from each sensor are then projected along the slit direction and combined as shown in Fig.~\ref{fig:plotccd}.


Each neutron is independent of the other. The neutron count at height $h$ above the mirror is $F(h) = \sum_i n_i P_i(h)$ where $n_i$ represents the neutron count in each state and $P_i(h)$ is the probability of detecting a neutron in state $\left|i\right>$ at height $h$. At the end of the mirror the neutron wave function is given by Eq.~\ref{eqn:schr}. Accounting the free fall over $x=1$\,mm at horizontal speed $v$, the propagated wave function reaching the detector is:
\begin{equation}
\tilde\psi_i(z,x)= \int_{-\infty}^{+\infty}\sqrt{\frac{mv}{2i\pi\hbar x}} e^{\frac{i}{\hbar}S(z,\zeta,x)}\psi_i(\zeta)\textrm{d}\zeta,
\end{equation}
where:
\begin{equation}
S(z,\zeta,x)=\frac{mv(z-\zeta)^2}{2x}-\frac{mgx}{2v}(z+\zeta)-\frac{mg^2x^3}{24v^3},
\end{equation}
is the classical free fall action after time $t=\frac{x}{v}$. The probability distribution is given by the square of the wave function integrated over the velocity distribution and convoluted with the detector resolution. The horizontal velocity distribution of the UCN is limited by the
Fermi potential of the source surface and is assumed uniform between $v_1=0.5$\,m.s$^{-1}$ and $v_2=4.6\,$m.s$^{-1}$:
\begin{equation}
P_i(h)= \alpha \int_{v_1}^{v_2}\int_{-\infty}^{+\infty}  \left|\tilde\psi_i(h,x)\right|^2e^{-\left(\frac{h-\eta}{2\sigma_d}\right)^2}\textrm{d}\eta\textrm{d}v,
\end{equation}
where the normalization factor $\alpha$ is the same for all templates and can be absorbed in the $n_i$ when fitting the data as only relative fractions will be of interest later on.

Three configuration were studied: the absorber slit set at $48\pm 1\,\mu$m with a step  $15\pm6\,\mu$m high (states $\left|1\right>$, $\left|2\right>$ and $\left|3\right>$ going through), and the absorber set $8 \pm 0.2\,\mu$m lower with and without the step (only states $\left|1\right>$ and $\left|2\right>$ going through). 
A global fit of the 3 data sets is performed, where the relative observed UCN flux and the measured absorber transmission (see section~\ref{sec:abs}) constrain the step and absorber heights (the difference between the two absorber height being fixed at $8\,\mu$m) and the shape of the distribution determines the contribution of each quantum state up to the sixth state. A global vertical offset is allowed to define the zero of the mirror. Data are compatible with an absorber height of $39\pm2~\mu$m and a step of $6\pm1\mu$m with a $p$-value of 0.33, in agreement with the settings. 

The fitted populations are given in table~\ref{tab-fitccd}. To explain the observed structure, higher energy states are needed than those authorized by the absorber transmission. We interpret them as excitations from the lower energy states between the absorber and the detector (either from scattering on the walls of the mirror, imperfection of the mirror itself or transition when going out of the absorber) that are not fully understood.  
The fit results are given in Fig.~\ref{fig:plotccd} where one can see the effect of absorber suppressing the third state (difference between the green and red curves), and of the step suppressing the first and part of the second state (difference between the blue and green curves) as well as the dip due to the node of the second state.

\begin{table}[!ht]
\begin{center}
\begin{small}
\caption[]{\label{tab-fitccd}
 Fitted step height, absorber height and population of each quantum states for the 3 configurations.}
\begin{tabular}{|cc||ccc|} \hline
Step ($\mu$m) & Abs. ($\mu$m) & $\left|1\right>$ & $\left|2\right>$ & $\left|3\right>$ to $\left|6\right>$ \\
\hline
$6\pm1$       & $47\pm 2$  & $7\pm 2\%$&	$24\pm 3\%$ &	$69\pm 8\%$  \\
$6\pm1$       & $39\pm 2$  & $15\pm 3\%$&	$28\pm 5\%$&	$58\pm 11\%$  \\
no step       & $39\pm 2$  & $37\pm 6\%$ &	$32\pm 8\%$&	$31\pm 15\%$  \\

\hline
\end{tabular}
\end{small}
\end{center}
\end{table}
\
\section{\label{sec:abs}Transmission of a rough absorber}

Higher energy states can be suppressed using a rough mirror placed on top of the propagation mirror top. The characteristic height at which a given quantum state is transmitted depends on the time the neutron spends inside the slit and and roughness of the surface. Following the conventions of~\cite{Escobar11}, the transmission of the rough absorber can be written as:
\begin{equation}
\sum_i A_i e^{\Phi b_i(h,z_0)}
\textrm{ with }
b_i = 5\times 10^{-4} z_0 \left|\psi_i^h\left(\frac{h}{z_0}\right)\right|^2.
\end{equation}
The characteristic size $z_0$ is included as a free parameter that will be fitted on the data. All the effects of the roughness as well as the time spend in the absorber slit is encoded in a single dimensionless parameter $\Phi$ whereas $b_i$ contains the presence probability of the neutron on the absorber surface. Quantum structures becomes visible in the transmission curve if $\Phi>40$.

Within a slit of height $u=\frac{h}{z_0}$, assuming the material of the slit has a Fermi potential $V$ much larger than the neutron vertical energies, the neutron wave functions on the top of the absorber are modified to:
\begin{equation}
\psi^h_i(u) = \left(\frac{mg}{V}\frac{K_i(u)^2}{K_i(0)^2-K_i(u)^2}\right)^\frac{1}{2},
\end{equation}
with
\begin{equation}
K_i(u) = Ai'\left(u-\lambda_i\right)-\frac{Ai\left(-\lambda_i\right)}{Bi\left(-\lambda_i\right)}Bi'\left(x-\lambda_i\right),
\end{equation}
where $E_0\lambda_i(u)$ is the energy of the $i$-th state, given by the $i$-th zero of  $Ai(u-\lambda_i)-Ai(u-\lambda_i)/Bi(u-\lambda_i)$.
This applies for a glass mirror where $V=84\,$neV and the quantized energies are of the order of a few peV. 

For the measurement, the absorber is placed at the end of the mirror and a $^3$He counter with a $15\,\mu$m titanium window is used~\cite{Roulier2015}. The titanium window has a negative Fermi potential, but due to the reflection on this potential and neutron capture within the foil the transmission efficiency is about $80\%$ at $v=4.6\,$m.s$^{-1}$ and drops to $40\%$ at $v=1.4\,$m.s$^{-1}$. The average velocity of detected neutrons is about $3\,$m.s$^{-1}$. This setup is shown in Fig.~\ref{fig:granitspectro}A. The absorber height is varied from $25\,\mu$m (contact) to $50\,\mu$m by $1\,\mu$m step and the UCN flux is recorded. To compensate for drifts in the input flux, the full range is scanned going up then down, down again and up over a few hours and the pattern is repeated several times. Measurement were done for several days over a period of 5 years leading to five independent data sets with UCN flux of 10 to 20\,mHz per state. For each $k=1\cdots5$, the neutron flux is modeled as:
\begin{equation}
T_k(h) = N_k + A_k \sum_i e^{\Phi b_i(h-O-r_k,z_0)},
\end{equation}
where we assume that all states are initially present in the same proportions and factorize the normalization factor $A_k$. The vertical position in all data sets is authorized to shift by a constant offset $O$ with respect to the slit height definition. We assume a Gaussian fluctuation with $\sigma_O=6\,\mu$m (the amplitude of the roughness). The relative position of each data set is also allowed a shift $r_k$  due to the reproducibility of the alignment procedure ($\sigma_r=1\,\mu$m).
The $k$-th data set consists of a series of measurement of UCN flux $f_{ik}$ with a Poisson statistical error $\Delta f_{ik}$ for different absorber heights $h_{ik}$. 
The constant background noise is measured at the same time as the data when the UCN source valve was closed. These measurements, varying from $2$ to $5$~mHz with a $5\%$ uncertainty, are used to constrain the $N_k$ parameters.
The five data sets are fitted simultaneously to the model. A likelihood function is build from a $\chi^2$ including the data and the constrains on parameters $O$, $r_k$ and $B_k$ according to their $1\,\sigma$ deviations.

\begin{figure}[!ht]
\includegraphics[width=0.45\textwidth]{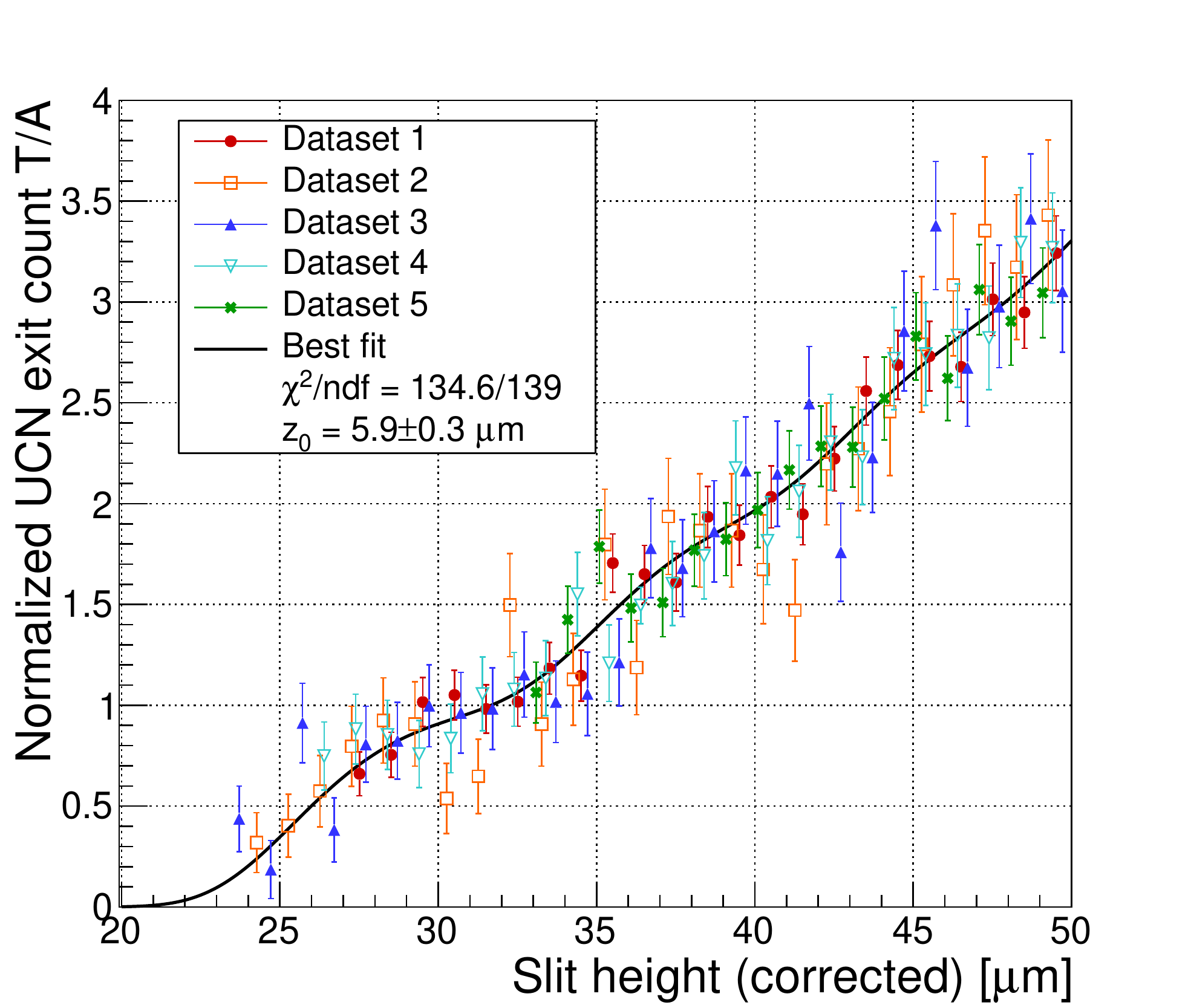}
\caption{\label{fig:absorber} Transmission of the absorber slit versus height. The background noise has been substracted and error bar enlarged accordingly. The best fit is presented as a blak curve.}
\end{figure}

All parameters, except $z_0$ are treated as nuisance parameters. We compute the profile likelihood $\tilde{\mathcal L}(z_0)$ by maximizing the likelihood function over all parameters for a fixed $z_0$. The fitted value of $z_0$ is obtained from the maximum of the profile likelihood and the error from the second derivative at maximum. Figure~\ref{fig:absorber} shows the normalized UCN count $(f_{ik}-B_k)/A_k$ as a function of the corrected slit height $h_{ik}-O-r_k$ for all data sets.
The $p$-value of the fit is 0.59, so the model describes correctly the data. The spatial extension of the wave functions is then measured as $z_0 = 5.9\pm 0.3\,\mu$m in agreement with the expected value. 

The $\Phi$ parameter of the GRANIT absorber was measured independently\cite{Escobar15} to $\Phi\approx 5000$ for a velocity of 4.5\,m.s$^{-1}$. As the velocity spectrum of the GRANIT source is centered around 3\,m.s$^{-1}$, we would expect $\Phi\approx7500$. A Monte-Carlo simulation of a tilted absorber showed that the lack of parallelism can drastically reduce the effective $\Phi$, especially a height difference between the entrance and output of the absorber slit. An average variation of of $6\pm0.5\,\mu m$ would reduce the absorber efficiency to the observed value $\Phi=109\pm 20$. This height is compatible with the precision of the alignment. Other effects, such as the polishing of the mirror and the gadolinium coating performed after the roughness measurements, could also contribute to the reduction of the efficiency. 

\section{Conclusion}
Using the GRANIT apparatus coupled to it superthermal $^4$He UCN source, we demonstrated the possibility to select gravitational quantum states of a bouncing neutron. This was done by using both a step to suppress low energy states and a rough absorbing slit to eliminate high energy states. We were able to measure the spatial extension of the wave function with a 5\% precision as $z_0 = 5.9\pm 0.3\,\mu\textrm{m}$. The absorber efficiency, quantified in the $\Phi$ parameter, is enough to reveal the expected step like quantum behavior.
The transmission of this absorber allows for the selection the first and second states. These methods could be further improved by ameliorating the mechanical setting, in particular the parallelism of the absorber slit. 

\begin{acknowledgments}
The authors would like to thank A.~Lacoste, Y.~Xi and A.~Bes at LPSC for the production on the conversion layer on CCD sensors, Y.~Carcagno, P.~Boge, E.~Perbet, L.~Viveargent, R.~Faure, O.~Bourrion and J.-P. Scordillis for their technical contributions that made the UCN source and the GRANIT spectrometer operational. LPSC Grenoble and ILL acknowledge the support of the French Agence Nationale de la Recherche (ANR) under Reference No. ANR-05-BLAN-0098. S.~Baessler gratefully acknowledge funding support from NSF through grant PHY-0855610.
\end{acknowledgments}

\bibliographystyle{ieeetr}

\begin{thebibliography}{10}

\bibitem{Lushchikov69}
V.~I. Lushchikov, Y.~N. Pokotilovskii, A.~V. Strelkov, and F.~L. Shapiro,
  ``Observation of ultracold neutrons,'' {\em JETP Lett.}, vol.~9, p.~23, 1969.

\bibitem{Steyerl69}
A.~Steyerl, ``Measurements of total cross sections for very slow neutrons with
  velocities from 100 m/s to 5 m/s,'' {\em Phys. Lett. B}, vol.~29, p.~33,
  1969.

\bibitem{Nesvizhevsky2000}
V.~V. Nesvizhevsky, H.~G. Börner, A.~M. Gagarski, G.~A. Petrov, A.~K.
  Petukhov, H.~Abele, S.~Bae\ss{}ler, T.~Stöferle, and S.~M. Soloviev,
  ``Search for quantum states of the neutron in a gravitational field:
  gravitational levels,'' {\em Nucl. Instrum. Methods}, vol.~A440, p.~754,
  2000.

\bibitem{Nesvizhevsky2002}
V.~V. Nesvizhevsky {\em et~al.}, ``Quantum states of neutrons in the earth's
  gravitational field,'' {\em Nature}, vol.~415, p.~297, 2002.

\bibitem{Jenke2011}
T.~Jenke, P.~Geltenbort, H.~Lemmel, and H.~Abele, ``Realization of a
  gravity-resonance-spectroscopy technique,'' {\em Nature Phys.}, vol.~7,
  p.~468, 2011.

\bibitem{Cronenberg2018}
G.~Cronenberg, P.~Brax, H.~Filter, P.~Geltenbort, T.~Jenke, G.~Pignol,
  M.~Pitschmann, M.~Thalhammer, and H.~Abele, ``Acoustic rabi oscillations
  between gravitational quantum states and impact on symmetron dark energy,''
  {\em Nature Phys.}, vol.~14, pp.~1022--1026, 2018.

\bibitem{Roulier2015}
D.~Roulier, F.~Vezzu, S.~Bae\ss{}ler, B.~Clément, D.~Morton, V.~V.
  Nesvizhevsky, G.~Pignol, and D.~Rebreyend, ``Status of the {GRANIT}
  facility,'' {\em Adv. High Energy Phys.}, vol.~2015, p.~730437, 2015.

\bibitem{Baessler2011}
S.~Bae\ss{}ler {\em et~al.}, ``New methodical developments for {GRANIT},'' {\em
  Comptes Rendus Physique}, vol.~12, p.~729, 2011.

\bibitem{Ichikawa14}
G.~Ichikawa, S.~Komamiya, Y.~Kamiya, Y.~Minami, M.~Tani, P.~Geltenbort, K.~Yamamura, M.~Nagano, T.~Sanuki, S.~Kawasaki, M.~Hino, M.~Kitaguchi, ``Observation of the spatial distribution of
  gravitationally bound quantum states of ultracold neutrons and its derivation
  using the wigner function,'' {\em Phys. Rev. Lett.}, vol.~112, p.~071101,
  2014.

\bibitem{Schmidt-Wellenburg09}
P.~Schmidt-Wellenburg {\em et~al.}, ``{Ultracold-neutron infrastructure for the
  gravitational spectrometer GRANIT},'' {\em Nucl. Instrum. Meth. A}, vol.~611,
  p.~267, 2009.

\bibitem{Zimmer11}
O.~Zimmer, F.~M. Piegsa, and S.~N. Ivanov, ``Superthermal source of ultracold
  neutrons for fundamental physics experiments,'' {\em Phys. Rev. Lett.},
  vol.~107, p.~134801, Sep 2011.

\bibitem{Golub75}
R.~Golub and J.~M. Pendlebury, ``The interaction of ultracold neutrons (ucns)
  with liquid helium and a superthermal ucn source,'' {\em Phys. Lett. A},
  vol.~53, p.~133, 1975.

\bibitem{Psw07}
P.~Schmidt-Wellenburg, J.~Barnard, P.~Geltenbort, V.~V. Nesvizhevsky,
  C.~Plonka, T.~Soldner, and O.~Zimmer, ``Efficient extraction of a collimated
  ultra-cold neutron beam using diffusive channels,'' {\em Nucl. Instr. Meth.
  A}, vol.~577, p.~623, 2007.

\bibitem{Barnard08}
J.~Barnard and V.~V. Nesvizhevsky, ``Analysis of a method for extracting
  angularly collimated ucns from a volume without loosing the density inside,''
  {\em Nucl. Instr. Meth. A}, vol.~591, p.~431, 2008.

\bibitem{Nesvizhevsky04}
V.~V. Nesvizhevsky, ``Investigation of neutron quantum states in the
  terrestrial gravitational field above a mirror,'' {\em Physics Uspekhi},
  vol.~47, p.~515, 2004.

\bibitem{Escobar15}
M.~Escobar, F.~Lamy, A.~E. Meyerovich, and V.~V. Nesvizhevsky, ``Rough mirror
  as a quantum state selector: Analysis and design,'' {\em Adv. High Energy
  Phys.}, vol.~2014, p.~8, 2015.

\bibitem{Voronin2006}
A.~Y. Voronin, H.~Abele, S.~Bae\ss{}ler, V.~V. Nesvizhevsky, A.~K. Petukhov,
  K.~V. Protasov, and A.~Westphal, ``Quantum motion of a neutron in a waveguide
  in the gravitational field,'' {\em Phys. Rev. D}, vol.~73, p.~044029, Feb
  2006.

\bibitem{Meyerovich06}
A.~E. Meyerovich and V.~V. Nesvizhevsky, ``Gravitational quantum states of
  neutrons in a rough waveguide,'' {\em Phys. Rev. A}, vol.~73, p.~063616, Jun
  2006.

\bibitem{Meyerovich07}
R.~Adhikari, Y.~Cheng, A.~E. Meyerovich, and V.~V. Nesvizhevsky, ``Quantum size
  effect and biased diffusion of gravitationally bound neutrons in a rough
  waveguide,'' {\em Phys. Rev. A}, vol.~75, p.~063613, Jun 2007.

\bibitem{Escobar11}
M.~Escobar and A.~E. Meyerovich, ``Beams of gravitationally bound ultracold
  neutrons in rough waveguides,'' {\em Phys. Rev. A}, vol.~83, p.~033618, Mar
  2011.

\bibitem{Bourrion2018}
O.~Bourrion, B.~Clément, D.~Tourres, G.~Pignol, Y.~Xi, D.~Rebreyend, and V.~V.
  Nesvizhevsky, ``{C2D8}: An eight channel {CCD} readout electronics dedicated
  to low energy neutron detection,'' {\em Nucl. Instrum. Methods}, vol.~A880,
  p.~28, 2018.

\bibitem{Clement19}
B.~Clément, A.~Bes, A.~Lacoste, R.~Combe, G.~Nesvizhevsky, Valery
  V.and~Pignol, Y.~Xi, and D.~Rebreyend, ``Boron-10 conversion layer for
  ultracold neutron detection,'' {\em JINST}, vol.~14, no.~09, p.~P09003, 2019.

\bibitem{Clement22}
B.~Clément, L.~Gesson, T.~Jenke, V.~V. Nesvizhevsky, G.~Pignol, S.~Roccia, and
  J.-P. Scordillis, ``Spatial resolution determination of a position sensitive
  ultra-cold neutron detector,'' {\em (unpublished)}, no.~arxiv 2203.09335,
  2022.

\end{thebibliography}

\end{document}